\documentclass[final,5p,times,twocolumn,authoryear]{elsarticle}

\usepackage{amsmath,bm}

\usepackage{amssymb}

\usepackage{enumitem}

\usepackage{color}

\newcommand{\degree}{\ensuremath{^\circ}}
\newcommand{\degrees}{\ensuremath{^\circ}}

\journal{Icarus}

\begin{document}

\begin{frontmatter}



\title{Equatorial locations of water on Mars: \\
Improved resolution maps based on Mars Odyssey Neutron Spectrometer data}

\author[1]{Jack T. Wilson\fnref{now}}
\fntext[now]{Present address: The Johns Hopkins Applied Physics Laboratory, 11100 Johns Hopkins Road, Laurel, MD 20723, USA}
\author[1]{Vincent R. Eke}
\author[1]{Richard J. Massey}
\author[2]{Richard C. Elphic}
\author[3]{William C. Feldman}
\author[4]{Sylvestre Maurice}
\author[5]{Lu\'{\i}s F. A. Teodoro}

\address[1]{Institute for Computational Cosmology, Department of Physics, Durham University, Science Laboratories, South Road, Durham DH1 3LE, UK}
\address[2]{Planetary Systems Branch, NASA Ames Research Center, MS 2453, Moffett Field, CA,94035-1000, USA}
\address[3]{Planetary Science Institute, Tucson, AZ 85719, USA}
\address[4]{IRAP-OMP, Toulouse, France}
\address[5]{BAER, Planetary Systems Branch, Space Sciences and Astrobiology Division, MS 245-3, NASA Ames Research Center, Moffett Field, CA 94035-1000, USA}

\begin{abstract}

We present a map of the near subsurface hydrogen distribution on Mars, based on epithermal neutron data from the Mars Odyssey Neutron Spectrometer.
The map's spatial resolution is approximately improved two-fold via a new form of the pixon image reconstruction technique.
We discover hydrogen-rich mineralogy far from the poles, including $\sim$10\,wt.~\% water equivalent hydrogen (WEH) on the flanks of the Tharsis Montes and $>$40\,wt.~\% WEH at the Medusae Fossae Formation (MFF). The high WEH abundance at the MFF implies the presence of bulk water ice.
This supports the hypothesis of recent periods of high orbital obliquity during which water ice was stable on the surface.
We find the young undivided channel system material in southern Elysium Planitia to be distinct from its surroundings and exceptionally dry; there is no evidence of hydration at the location in Elysium Planitia suggested to contain a buried water ice sea.
Finally, we find that the sites of recurring slope lineae (RSL) do not correlate with subsurface hydration.
This implies that RSL are not fed by large, near-subsurface aquifers, but are instead the result of either small ($<$120\,km diameter) aquifers, deliquescence of perchlorate and chlorate salts or dry, granular flows.

\end{abstract}


\end{frontmatter}


\section{Introduction}\label{sec:introduction}

The main goal of the Mars Odyssey Neutron Spectrometer (MONS) is to determine the major near-surface reservoirs of hydrogen on Mars \citep{Feldman2004}. Knowing the present distribution of water in the Martian near-subsurface is important for several reasons: it allows inferences about the past and present climate to be drawn, which, in turn, give information about the dynamic history of Mars and the possibility of the past, or present, existence of life. Additionally, understanding the small-scale distribution of water is important for landing site selection for missions looking for signs of life or exploring in-situ resource utilisation \citep{VV}. MONS data have been used to map the hydrogen content of the Martian near-subsurface \citep{Feldman2002Sci,Tokar2002,Feldman2004} on $\sim 550$~km scales \citep{Teodoro2013}.  Hydrogen rich deposits, with WEH content $>$ 25~wt.~\%,  were found poleward of $\pm$60\degrees, which were interpreted as water-ice buried under a layer of desiccated soil \citep{Feldman2002Sci}.  Additional low-latitude hydrogen deposits were observed at Arabia Terra and Elysium Planitia \citep{Feldman2002Sci} with 9.5$\pm$1.5~wt.~\% WEH \citep{Feldman2004GRL}. Water ice should not be stable equatorward of $\pm$ 30\degrees\ \citep{Mellon93}, which has led to the suggestion that these equatorial hydrogen deposits are in the form of hydrated minerals \citep{Feldman2002Sci}.

The $\sim 550$ km spatial resolution of the MONS instrument suppresses smaller-scale features in the MONS data, and the inferred hydrogen distribution. It also results in a reduction in the dynamic range of the data, leading to an underestimate in the wt.~\%~WEH content of small hydrogen-rich areas. The previously inferred WEH abundances for equatorial features will have been underestimated because of this effect. In this paper we will develop and apply an image reconstruction technique based on the pixon method, which has been used to successfully reconstruct planetary data \citep{Lawrence2007,Elphic2007,Eke2009,Wilson2015}, to improve the resolution of the global MONS data set in a way that is robust to noise.  This will be the first application of a Bayesian image reconstruction technique to a globally defined remotely-sensed planetary data set.

We will focus on a few locations that have been proposed to contain water in the equatorial regions of Mars deposited in the geologically recent past.  This water is hypothesised to have been deposited during past periods of high orbital obliquity when the water ice currently at the poles becomes unstable and is ultimately deposited elsewhere \citep{Forget2006}. Evidence for equatorial hydration is both morphological \citep{Head2014} and compositional \citep{Feldman2004}, and is seen at both the Medusae Fossae Formation  (MFF) and the Tharsis Montes.

The MFF is a discontinuous geological unit of easily erodible material that stretches $\sim$ 1000\,km across equatorial latitudes, along the boundary of the northern lowlands and southern highlands, located in both Elysium and Amazonis Planitiae.  The origin of the MFF is uncertain. Proposed explanations include consolidated pyroclastic deposits \citep{Scott82} and aeolian sediments with ice-rich material \citep{Head2004}, similar to that found in the polar layered deposits \citep{Schultz88}, laid down during periods of high orbital obliquity. Radar sounding using the MARSIS instrument onboard ESA's Mars Express has been used to measure the dielectric constant of the MFF material and found it to be consistent with the MFF containing a large component of water ice or anomalously low density soil \citep{Watters2007}. It may be possible to distinguish between these two mechanisms using neutron derived hydrogen abundances. 

Evidence, in the form of surface morphology and cratering, for late Amazonian tropical mountain glaciers on the north-western flanks of the Tharsis Montes and Olympus Mons has been detailed extensively using observations from the Mars Express, Mars Global Surveyor and Mars Odyssey orbiters \citep{Head2003,Shean2005,Shean2007}. The production of such glaciers at equatorial Mars today is impossible given the  current climatic conditions. Thus, the existence of these glaciers is interpreted to be the result of the migration of volatiles (chiefly water) from the poles to the equator during past periods of high orbital obliquity. Climate models predict the accumulation of ice on the north-western slopes of these mountains, during hypothesised periods of high obliquity, due to the adiabatic cooling of moist polar air \citep{Forget2006,Madeleine2009}.  

The extent to which these equatorial deposits remain and the form in which they are present are not yet settled questions. \citet{Campbell2013} see no evidence for buried ice at Pavonis Mons using the Shallow Radar (SHARAD) instrument onboard the Mars Reconnaissance Orbiter. However, morphological evidence is presented by \citet{Head2014} in the form of fresh ring-mold craters at both Pavonis and Arsia Mons, which is suggestive of the presence of buried ice today or in the very recent past.

The utility of this data set in constraining the distribution of water on the scale of the tropical mountain glaciers has been limited by its relatively poor spatial resolution due to the $\sim$ 550\,km full width at half maximum, FWHM, footprint of the MONS.  Jansson's method was used by \citet{Elphic2005} to perform image reconstruction on the MONS data at Tharsis. However, it is known that this method amplifies noise and may introduce spurious features \citep{Prettyman2009}. 

We will use our improved resolution map of the epithermal neutron count rate across the entire surface of Mars to look for evidence of buried, non-polar hydrogen in the form of both hydrated minerals and water ice at the Tharsis Montes and the Medusae Fossae Formation.  In addition, we will examine several non-polar sites where water is suggested to be present in some form: Meridiani Planum, Elysium Planitia and the southern-mid latitudes, where RSL are observed. The increase in resolution over the unreconstructed MONS data allows better correspondence between structures in the neutron data and features in surface imagery, thus enabling a more robust geophysical interpretation.

In the next section we describe the modifications made to the pixon method and the new algorithm that has been developed.  In section~\ref{sec:data} we discuss the MONS data used in this study, along with the properties of the instrument, before presenting the results of the reconstructions in section~\ref{sec:results}.  Small scale features that emerge in the reconstructions are described in sections~\ref{sec:small}~and~\ref{sec:other}. Finally, we conclude in section~\ref{sec:conc}.

\section{Methods}\label{sec:meth}
In this section we will describe a new version of the pixon image reconstruction technique, adapted for use on the sphere.  The pixon method is a Bayesian image reconstruction technique, in which pixels are grouped together to form `pixons' \citep{PP93,Eke2001}. It is motivated by considering the posterior probability 
\begin{equation}\label{eqn:prob}
P(\hat{I},M|D) = \frac{P(D|\hat{I},M)P(\hat{I}|M)P(M)}{P(D)},
\end{equation}
where $\hat{I}$ is the reconstructed image and $M$ is the model, which
describes the relationship between $\hat{I}$ and the data, including
the point spread function, PSF, and the basis in which the image is represented. As the data are already
taken, $P(D)$ is not affected by anything we can do and is therefore
constant. Additionally, to avoid bias, $P(M)$ is assumed to be uniform. This assumption
leaves two terms: the first, $P(D|\hat{I},M)$, is the likelihood of
the data given a particular reconstruction and model, calculated using a misfit statistic. The second term, $P(\hat{I}|M)$, is the image prior. The grouping of pixels into pixons greatly reduces the number of degrees of freedom in the reconstructed image, improving the image prior.  The most likely reconstruction is that containing fewest pixons that is still able to provide a sufficiently good fit to the data.

\subsection{Locally adaptive pixon reconstruction}\label{sec:Pix}
Previous versions of the pixon method imposed a strict maximum entropy constraint on the
solution \citep{PP93,Eke2001}, in order to simplify the maximization of the posterior by
reducing the parameter space that is to be explored.  Here we make no such requirement, which will allow the creation of previously inaccessible
and more likely reconstructions.

As in the maximum entropy pixon method the reconstructed count-rate map, $\hat{I}$, is constructed by convolving a `pseudoimage',
$H$, with a Gaussian kernel, $K$, whose width, $\delta$,
may vary across the image, i.e.
\begin{equation}
\hat{I}({\bf x}) = (K_{\delta({\bf x})}*H)({\bf x}).
\end{equation}
Initially the smoothing scale, or pixon width, $\delta({\bf x})$, is large. The final value is determined
for each pixel using an iterative procedure, which includes the calculation of a newly defined local misfit statistic. In practice a finite set of distinct pixon sizes is used as this keeps
the time of a single calculation of the global misfit statistic, $E_R$, down to
$\mathcal{O}({n_{\mathrm{pixels}}\log{n_{\mathrm{pixels}}}})$ (i.e. that of a
fast Fourier transform), where $n_{\mathrm{pixels}}$ is the number of
pixels in the image, whereas if the pixon size were allowed to vary
continuously then the time would be
$\mathcal{O}({n_{\mathrm{pixels}}^2\log{n_{\mathrm{pixels}}}})$. In order to
generate $\hat{I}$ we interpolate linearly between the images based on
the two pixon sizes closest to those required for each pixel.

The following subsection describes the structure of the locally adaptive pixon algorithm before the details of the misfit statistic and prior calculation are detailed.

\subsubsection{Structure of the locally adaptive pixon algorithm}\label{sec:struc}

\begin{figure}
	\begin{center}
		\includegraphics[width=0.7\columnwidth]{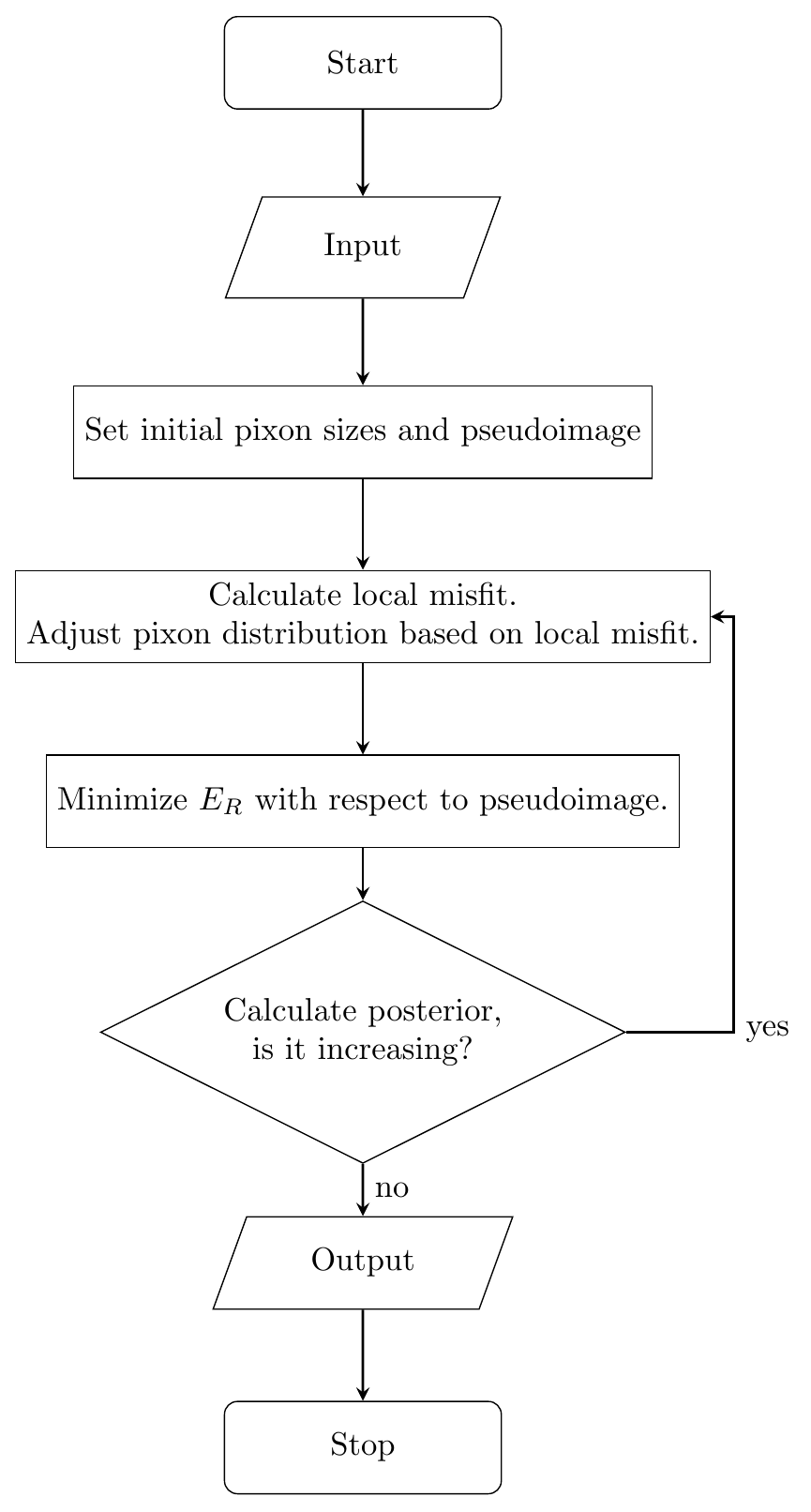}
		\end{center}
	\caption{A flow chart showing the locally adaptive pixon image reconstruction algorithm.}\label{fig:flow}
\end{figure}

Our goal is to maximize the posterior probability (equation~(\ref{eqn:prob})) whilst ensuring that the
residual field should be indistinguishable from a random Gaussian field.  We do this using an iterative
procedure with two stages in each iteration (figure~\ref{fig:flow}). Firstly, for a
given pixon size distribution the
values in the pseudoimage are adjusted to minimize the global misfit statistic (see section~\ref{ssec:E_R}) using the Polak-Ribi\`ere
conjugate gradient minimization algorithm
\citep{NR}. Secondly, we calculate the value of the posterior and, for each pixel in the data-grid, the local misfit statistic (section~\ref{ssec:LocalE_R}). These local misfit statistics are then ordered to form their cumulative distribution function (CDF).  The
difference, in each pixel, between the actual and expected, $\chi^2$-distributed, CDF is then used to modify the pixon
size in the same pixel.  The change in pixon width is directly proportional to this difference, with the
constant of proportionality determined empirically such that structure appears in the image sufficiently
slowly to avoid introducing spurious features.

The combination of global misfit statistic minimization and pixon size modification is repeated, starting with large pixons and consequently a flat, constant reconstructed image everywhere, until the posterior probability stops increasing.

\subsubsection{The global misfit statistic, $E_R$}\label{ssec:E_R}
The misfit statistic is derived from the reduced residuals between the
data and the blurred model, 
\begin{equation}\label{eqn:res}
R({\bf x}) = \frac{D({\bf x}) - (\hat{I}*B)({\bf x})}{\sigma({\bf x})},
\end{equation}
where $\sigma({\bf x})$ is the anticipated statistical noise in pixel
${\bf x}$.
Rather than using $\chi^2=\sum R^2$, we adopt $E_R$ from \citet{PP92} as the
misfit statistic. This statistic is defined as
\begin{equation}\label{eqn:ER}
E_R = \sum_{{\bf y}}A_{R}({\bf y})^{2},
\end{equation}
where $A_R$ is the autocorrelation of the residuals, $A_R({\bf y})
= \int R({\bf x}) R({\bf x} + {\bf y}){\rm d}{\bf x}$ for a 2-dimensional pixel separation, or lag, of ${\bf y}$. 
The benefit of minimizing $E_R$, over the more
conventional $\chi^2$, is that doing so suppresses spatial
correlations in the residuals, preventing spurious features
being formed by the reconstruction process. \citet{PP92}
recommend that the autocorrelation terms defining $E_R$ should be those corresponding to pixel separations smaller than 
the instrumental PSF. For a well-sampled PSF this means
many different pixel 
lags. However, for the reconstructions we have attempted,
there is negligible
difference between those including different
numbers of pixel lags. We therefore use only terms
with adjacent pixels (including diagonally adjacent pixels in 2D) to speed up the
computation.  Although each pixel is adjacent to eight others the symmetry implies that $A_R({\bf y}) = A_R(-{\bf y})$, so only the four distinct terms need be included when calculating $E_R$.

Prior to this work the pixon method has been implemented only on flat planes.  When calculating $E_R$ on a spherical surface we decided to map the sphere to a flat (i.e. Gaussian curvature equal to zero) torus.  This is done via an equirectangular projection of the sphere to a square Cartesian plane (by extending the regular parameterization of the sphere to $0<\theta<2\pi$, $0<\phi<2\pi$) and identifying opposite edges.  This mapping leaves the details of the calculation, including the use of FFTs for fast convolution, unchanged from the flat 2D case. However, as the lag terms included in $E_R$ cover a certain number of pixels this mapping has the side effect of mixing the physical scales over which the statistic is defined.

\subsubsection{Local misfit statistic, $E^\prime_R$}\label{ssec:LocalE_R}
To create a local $E_R$ statistic, ${E_R^\prime}$, we begin by localizing $A_R$ by multiplying with a kernel, $K$, such that

\begin{equation}
A^\prime_R({\bf x},{\bf z}) = \int R({\bf y}) R({\bf x} + {\bf y}) K({\bf z} - {\bf y}) \, d{\bf y},
\end{equation}
which we can rewrite in a form resembling a convolution, by defining $C_{\bf x}({\bf y}) = R({\bf 
y})R({\bf x} + {\bf y})$, as

\begin{equation}\label{eqn:A_R_prime}
A^\prime_R({\bf x},{\bf z}) = \int C_{\bf x}({\bf y}) K({\bf z} - {\bf y}) \, d{\bf y}.
\end{equation}
We are free to choose the kernel and for ease of calculation select a gaussian, the width of which should be larger than the largest pixon width so that it is the pixon size, and not its gradient, that sets the smoothing scale in the reconstruction. For use in determining pixon widths we need to know the distribution of $A^\prime_R$.  Noting that $C_{\bf 
x}({\bf y})$ has a normal product distribution with width one and applying the central limit theorem 
allows one to deduce that $A^\prime_R$ must have a standard normal distribution so ${E_R^\prime}$ obeys a 
$\chi^2$ distribution.

When convolving in angular space we should define (using equation~(\ref{eqn:dxdohm})).
\begin{equation}
A^\prime_R({\bf x},\omega({\bf z})) = \left(\frac{C_{\bf x}}{a}*K\right)(\omega({\bf z})),
\end{equation}
and
\begin{equation}\label{eq:F_R}
{E_R^\prime}({\bf z}) = \sum_{{\bf y}}\frac{{A^\prime_R}^2({\bf y}, {\bf z})}{\sigma_{{A^\prime_R}}^2({\bf z})},
\end{equation}
where $\sigma_{A^\prime_R}^2$ can be deduced by analogy with a random walk i.e.
\begin{equation}
\begin{split}	
\sigma_{A^\prime_R}^2({\bf z}) &= \int K^2({\bf z} - {\bf y}) \, d{\bf y}\\
&= \left(\frac{1}{a}*K^2\right)(\omega({\bf z})).
\end{split}
\end{equation}

\subsubsection{Prior calculation}\label{ssec:Prior}
Counting arguments show that, in the pixon method, the prior is given by
\begin{equation}
\ln(P(I|M)) \approx N\ln\left(\frac{N}{n}\right) + \frac{1}{2}\ln(N) - \frac{n-1}{2}\ln(2\pi) - \sum_{i=1}^{n}\left(N_i + \frac{1}{2}\right)\ln(N_i),
\end{equation}
where $N$ is total amount of information in the image, $N_i$ the information content of pixon $i$ and $n$ is the number of pixons (see e.g. \citet{Eke2001} for a full derivation).  In the maximum entropy pixon method the content of each pixon was defined to be equal, so its explicit numerical calculation is never necessary (i.e. $N_i = N/n$).  However, our new pixons are not only inhomogeneous but there needn't be an integer number, with pixon number defined as
\begin{equation}
	n = \sum \limits_{\bf x} \frac{1}{2\pi\delta({\bf x})^2},
\end{equation}
i.e. the number of independent degrees of freedom in the image. For the case of a non-integer number of pixons we choose to weight the contribution of the final non-integer pixon, 
such that the natural logarithm of the prior (using Stirling's approximation) becomes
\begin{equation}
\begin{split}
\ln(P&(I|M)) \approx N\ln\left(\frac{N}{n}\right) + \frac{1}{2}\ln(N) - \frac{n-1}{2}\ln(2\pi) \\
&- \sum_{i=1}^{\lfloor n \rfloor}\left(N_i + \frac{1}{2}\right)\ln(N_i) - (n - \lfloor n \rfloor) \left(\left(N_n + \frac{1}{2} \right)\ln(N_n)\right),
\end{split}
\end{equation}
where $N_n$ is the information in the final fraction of a pixon scaled up to the size it would be if it 
were filling a whole pixon.

\section{Data}\label{sec:data}

Approximately 3.5 Martian years of the time-series MONS prism-1 \citep{Maurice2011} 
observations from JD 2452324.125388 (just before the start of Mars Year 26) until JD 2454922.577436 are 
used in this study. Prism-1 is the nadir facing detector on MONS and is most sensitive to epithermal 
neutrons.  In this paper we restrict our attention to the `frost-free' data, for which we use the same definition as \citet{Maurice2011} i.e. those data taken at a location (in time and space) in which the CO$_{\rm 2}$ area density in the NASA Ames GCM is less than 0.2 gcm$^{\rm -1}$.

As Mars Odyssey is in a near-polar orbit we chose to use an equirectangular pixelization, which results in approximately the same number of observations in each pixel.  The density of the pixelation is chosen so that the PSF is well sampled. In this work we use a square, 0.5\degree$\times$0.5\degree, grid of pixels in a equirectangular projection.  The statistical errors within a pixel are determined, empirically, by calculating the scatter between repeat observations in the observed count rate within the pixel.  It was necessary to estimate the noise amplitude per pixel empirically as the various corrections applied to the data, for atmospheric thickness, cosmic ray flux, and spacecraft latitude and altitude, resulted in the data no longer possessing Poisson statistics \citep{Maurice2011}.

\subsection{Assumed instrumental properties}\label{ssec:Assump}
\begin{figure}[]
	\begin{center}
	\includegraphics[trim=2cm 0cm 10cm
2cm,clip,width=\columnwidth]{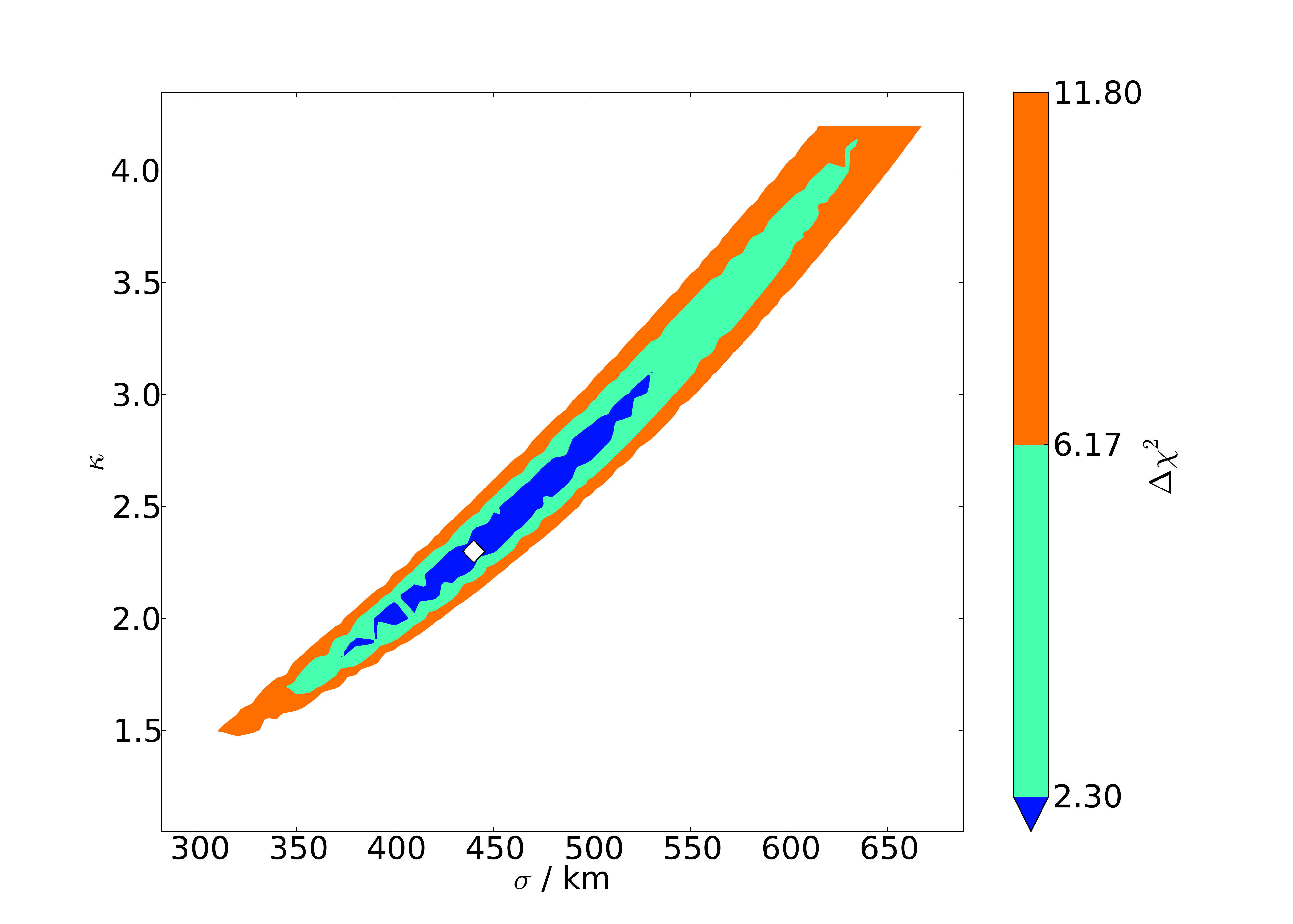}
	\end{center}
\caption{Constraints on the parameters in the PSF. The parameters are those in equation~\ref{eqn:kappa}.
The white diamond shows the best-fit reconstruction with minimum $\chi^2$. Coloured regions enclose probabilities of 68\%/95\%/99.7\%, determined using $\Delta\chi^2=\chi^2-\chi^2_\mathrm{min}$.}\label{fig:params}
\end{figure}

As a starting point for our estimation of the MONS PSF we have taken the calculated prism-1 spatial response of the MONS instrument to a circular feature with 5 degree diameter from 
\citet{Prettyman2009}.  This calculation of the response involved the creation of a detailed numerical model of the Martian surface and atmosphere and the Mars Odyssey spacecraft. The production of neutrons via cosmic ray interaction with the Martian atmosphere and surface and their subsequent transport within the surface, atmosphere and exosphere and detection within the spacecraft was performed using the Monte Carlo N-Particle eXtended (MCNPX) transport code.  To convert this `disk spread function' to the instrumental PSF, we calculated 
the result of convolving a $5\degree$ diameter disk with various trial PSFs, assuming that the PSF is azimuthally symmetric about the spacecraft's nadir point.  The profile resulting from these convolutions that 
most closely matches that obtained from the simulations is taken to be the correct PSF.  We make the 
assumption that the PSF can be described by a kappa function, defined as
\begin{equation}\label{eqn:kappa}
\frac{f(s)}{f(0)} = \left(1 + \frac{s^2}{\sigma^2}\right)^{-\kappa},
\end{equation}
where $\kappa$ and $\sigma$ are parameters to be determined, $s$ is the arc length from the sub-orbital 
point and $f(s)$ is the response of the instrument to a source at $s$. $\kappa$ functions are often used 
to model the response of spacecraft-borne gamma ray and neutron detectors \citep{Maurice2004}. 
The results of this calculation are shown in Fig.~\ref{fig:params}, from which the best fitting PSF has 
$\sigma = 440$ km and $\kappa = 2.3$, giving a 
FWHM of 520km. 

\section{Results}\label{sec:results}

\begin{figure*}[]
	\begin{center}
	\includegraphics[width=\textwidth]{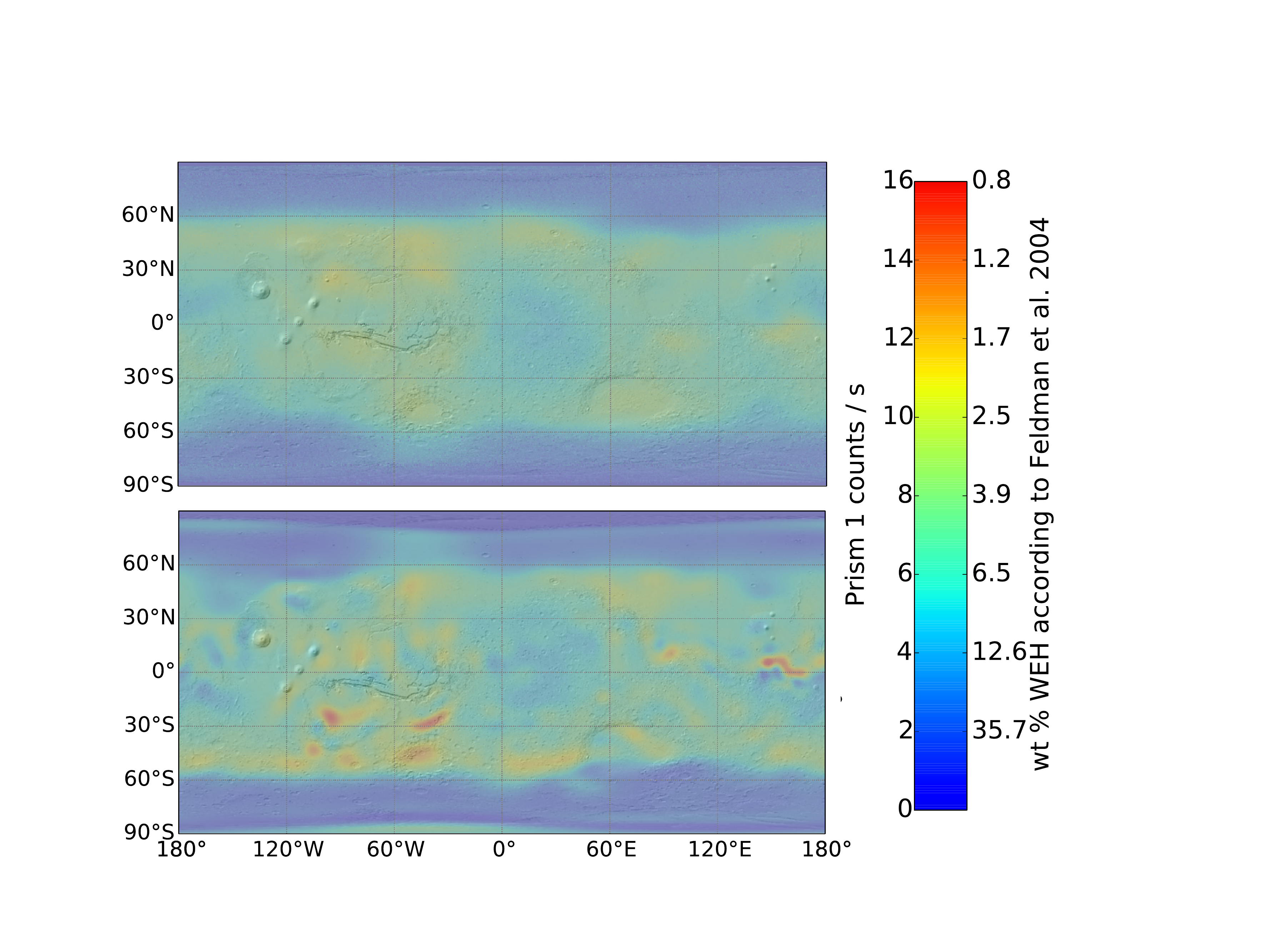}
	\end{center}
\caption{{\it Top.} The global frost-free MONS prism-1 data, used in the reconstruction. {\it Bottom.} A global locally-adaptive pixon reconstruction of the MONS prism-1 data.  Underlaid on both panels is a MOLA shaded relief map. The colour bar also shows the conversion to wt. \% water equivalent hydrogen from \citet{Feldman2004}.}\label{fig:global}
\end{figure*}

The global frost-free MONS data and our reconstruction of the underlying field are shown in the two panels of figure~\ref{fig:global}.  Locally adaptive pixon reconstruction of the data leads to an increase in dynamic range of the count rate of nearly 50\%, from $\sim$1-11 to $\sim$0-16 counts per second.  In addition to the improvement in dynamic range, there is an enhancement in spatial resolution, quantified below. All of the conversions from epithermal neutron count rate to weight \% water equivalent hydrogen (wt. \% WEH) in this work use the single layer model in \citet{Feldman2004}, i.e. the surface is assumed to be made up of a semi-infinite layer with a constant wt. \% WEH content. 

\subsection{Improvement in resolution}\label{sec:res}

To estimate the resolution of the reconstructed data we consider the two-point correlation function of the reconstruction. 
Comparing the separation required for the correlation function to drop to a given level allows the effective resolution of the 
reconstruction to be calculated. The correlation between the count rate at two points (with angular separation $\theta$) on the 
surface of the planet can be calculated using the relation

\begin{equation}\label{eqn:2pt}
	C(\theta) = \frac{1}{4\pi}\sum^{\infty}_{l=0}(2l+1)C_lP_l\cos\theta,
\end{equation}
where $P_l$ are the Legendre polynomials and $C_l$ is the power spectrum of the count rate, defined by

\begin{equation}
	C_l = \frac{1}{2l+1}\sum^{l}_{m=-l}|a_{lm}|^2,
\end{equation}
with $a_{lm}$ being the spherical harmonic coefficients defining the count rate field.

\begin{figure}[]
	\begin{center}
	\includegraphics[width=\columnwidth]{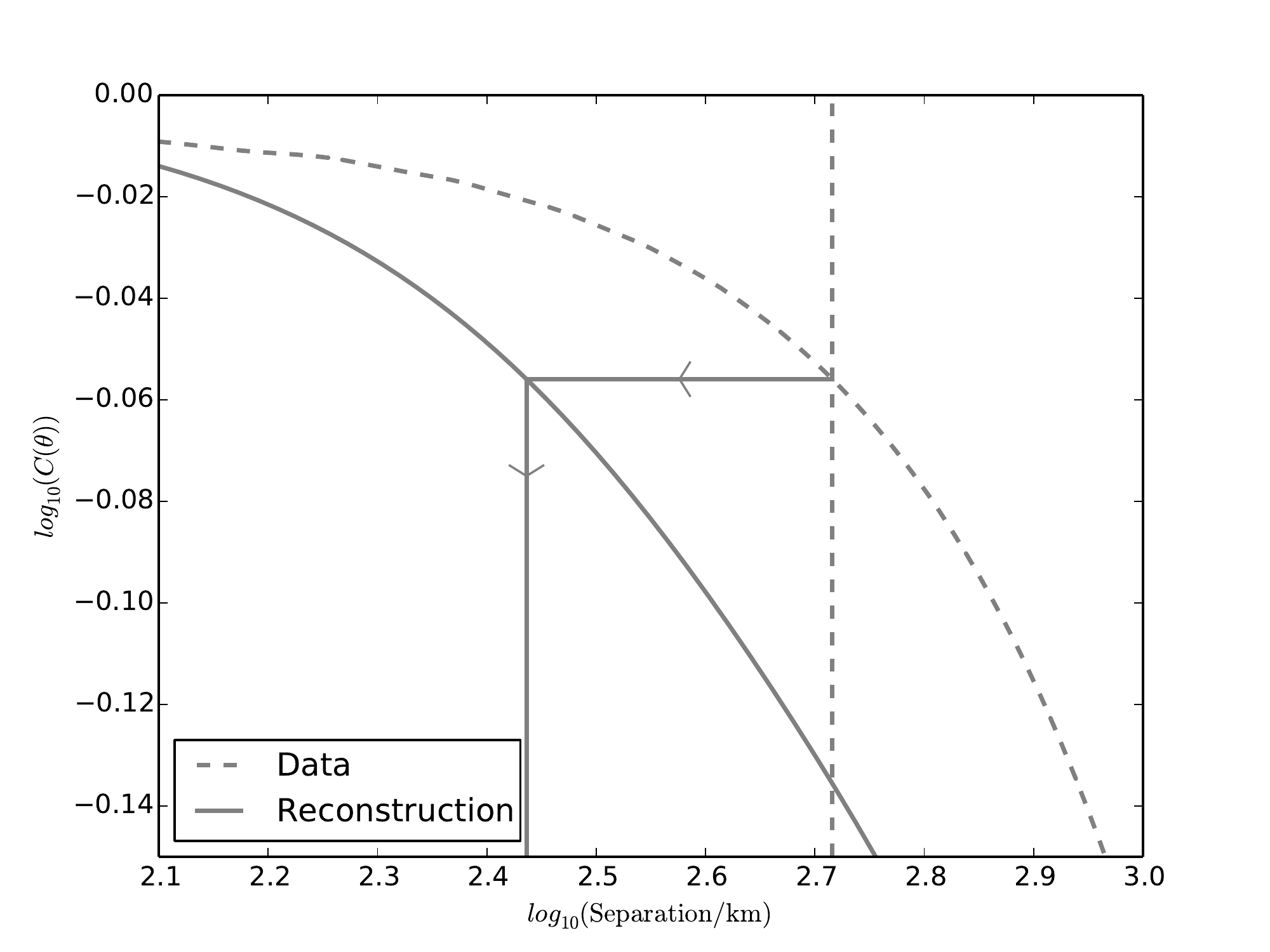}
	\end{center}
\caption{The two point angular correlation function of the MONS data and the 
reconstruction, normalized to unity at zero separation.  The vertical dashed line shows the MONS PSF FWHM 
scale.  The solid vertical line at log$_{\rm 10}$(s) = 2.46~(or  290 km) shows the separation required for the power in the reconstruction to drop by the same proportion that the power in the data drops at a separation equivalent to the FWHM of 
the PSF.}\label{fig:2pt}
\end{figure}

This function is plotted in figure~\ref{fig:2pt} for both the data and the reconstructed image. A separation equal to the FWHM of the data (520 km or log(separation) = 2.72)  corresponds to a normalized correlation value of log(C(${\rm theta}$)) = -0.06.  The reconstructed image reaches this normalized correlation value at a separation of 290 km (or log(separation) = 2.46). Thus the effective FWHM of the map is 290 km rather than 520 km i.e. variations on scales smaller than 290 km are unlikely to be found in our reconstruction. This three fold decrease in effective detector footprint leads to the enhanced dynamic range shown in the reconstruction in figure~\ref{fig:global}.  The value of $C(0)$ i.e. the power on the smallest scales, increases by 40\% between the data and reconstruction.

\subsection{Estimation of errors in the pixon method}\label{sec:errors}

\begin{figure}[]
	\begin{center}
	\includegraphics[width=\columnwidth]{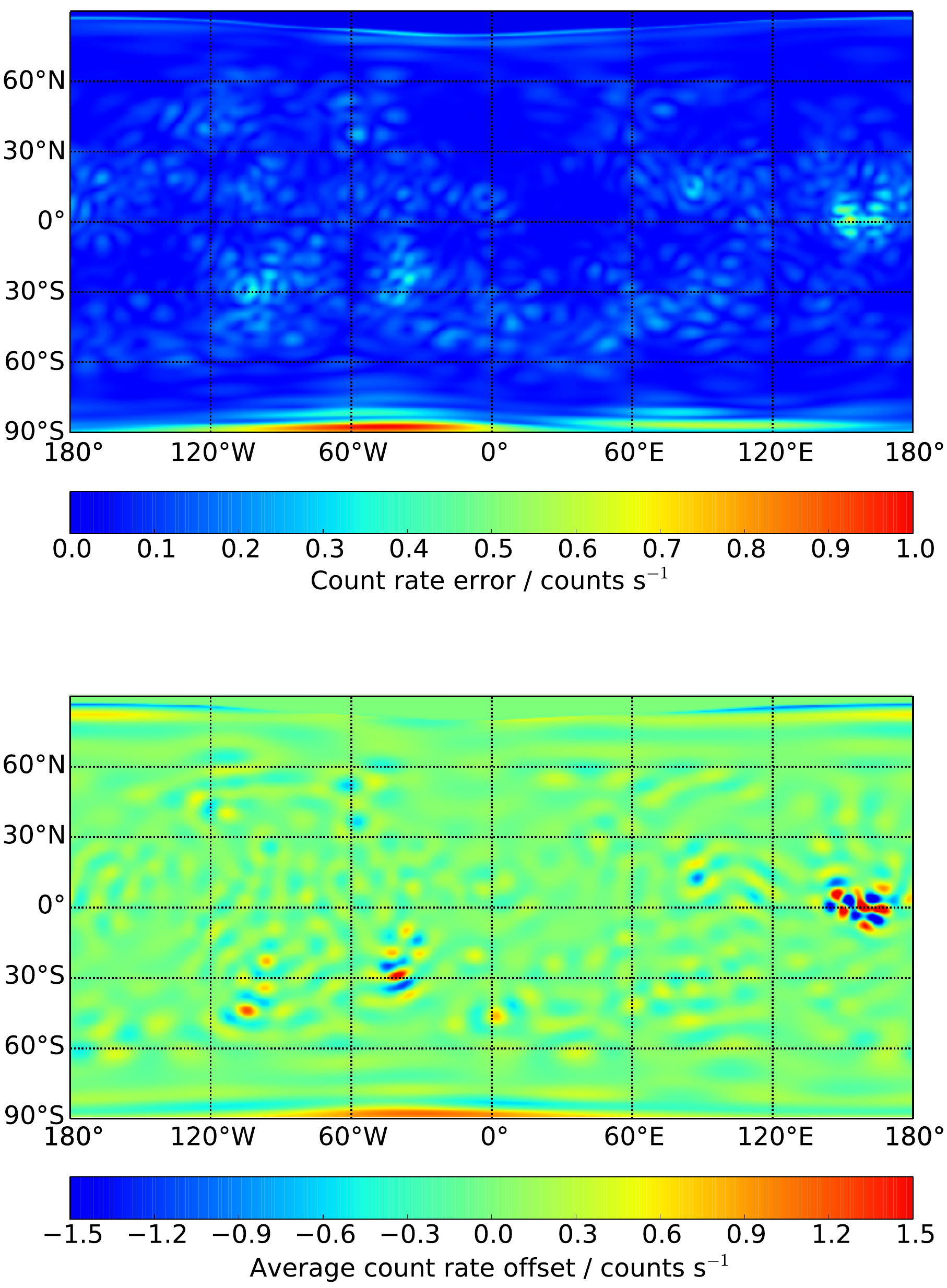}
	\end{center}
\caption{{\it Top.} The standard deviation of the reconstructed count rate based on 10 mock data sets. {\it Bottom.} A plot of the average difference, in each pixel, between the mock reconstructions and the true image.}\label{fig:datarec}
\end{figure}

To estimate the systematic and random errors in the pixon method, when applied to the MONS data, we created a set of 10 mock 
data sets.  These mock data were generated by blurring the reconstructed image (figure~\ref{fig:global} lower panel) with the PSF and then adding Gaussian noise with $\sigma$ as determined for the real data. Running the pixon reconstruction algorithm on each of these mock data sets enables the errors in the reconstructed count rate, both statistical and systematic, to be estimated.  
The upper panel of Figure~\ref{fig:datarec} shows the RMS scatter in each pixel over the 10 reconstructions, a measure of the statistical error in the reconstruction due to noise in the data. Comparison with figure~\ref{fig:global} shows the noise to be greatest where the count rate is large or varies quickly, such as at the south polar cap.  The global mean statistical error in the reconstructed epithermal 
count rate is 0.26 counts per second.  The average difference between a reconstruction based on mock data from one based on the real data is shown in figure~\ref{fig:datarec} (lower panel).  This provides a measure of the systematic errors in the reconstruction method and has a global average of 0.47 counts per second.  The regions with highest offset are mostly co-located with those with the largest random errors.  The feature at around 0\degrees\,N, 150\degrees\,E is particularly discrepant as it features regions of very low and high count rate in close proximity.  The inability to completely remove the smoothing effect of convolving with the PSF causes these regions to blend together.  The mottling in these images is an artefact related to the size of the pixons.

\subsubsection{Limit of detection of increased hydration}\label{sec:min_size}
\begin{figure}[]
	\begin{center}
	\includegraphics[width=\columnwidth]{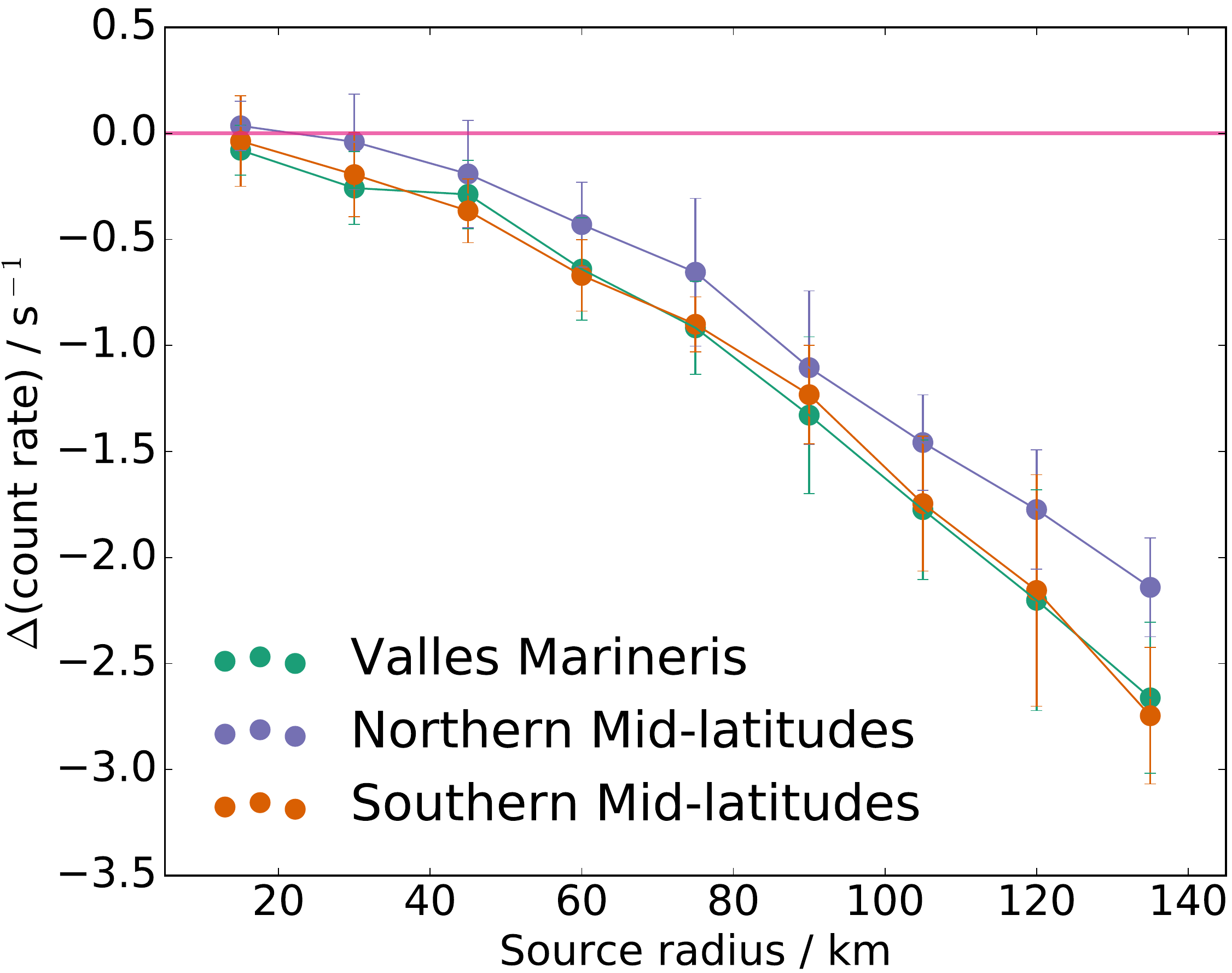}
	\end{center}
\caption{The difference in count rate between reconstructions with 100 wt\% WEH sources and those without as a function of source radius.  1$\sigma$ error bars are calculated from the standard deviation of 10 reconstructions with different noise realizations. The colours correspond to the regions where the sources were inserted. The horizontal line is at count rate difference equal to zero.}\label{fig:rsl_count}
\end{figure}

To place a limit on the smallest source region that could be detected using the new pixon method we created a series of mock data sets containing circular regions of 100 wt\% WEH with various diameters, located in some regions considered in this work, i.e. the southern mid-latitudes, Valles Marineris and the northern mid-latitudes.  We then performed reconstructions of these mock data sets and examined the reconstructions to see how large the 100 wt\% source region must be before a statistically significant depression in the count rate is seen.  The variation of count rate with source size is shown in figure~\ref{fig:rsl_count}, where the error bars show the scatter in a set of ten mock data sets. It can be seen that sources smaller than approximately 60 km in radius do not lead to statistically significant decreases in count rate so are not reliably resolved using this method.  Therefore, we cannot detect near-subsurface water sources smaller than 120 km in diameter.

\subsection{Comparing reconstructions with surface measurements}
Our reconstructions also enable us to examine the region around Gale crater, the landing site and science target of the Curiosity rover.  The Curiosity rover has an active neutron sensor onboard, the Dynamical
Albedo of Neutrons (DAN) instrument  \citep{Mitrofanov2012}.  Unlike for passive, orbital remote sensing, DAN works by illuminating the surface
with a pulsed neutron source and then measuring the returned neutron flux to infer composition.  This gives us a ground truth against which to compare our reconstructions.

\begin{figure*}[]
	\begin{center}
	\includegraphics[width=\textwidth]{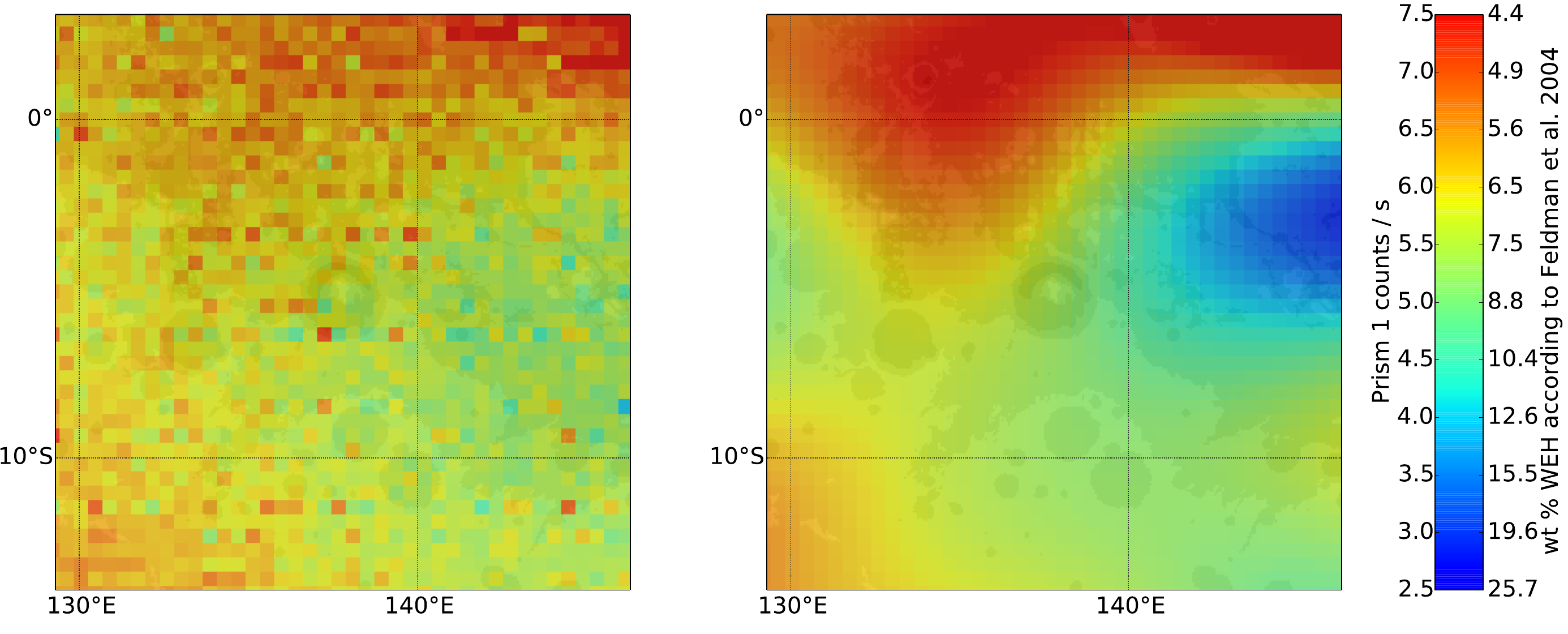}
	\end{center}
\caption{ The MONS data (left) and reconstruction (right) in the area around Gale Crater. Underlaid is a MOLA shaded relief map. The colour bars also show the conversion to wt. \% water equivalent hydrogen from \citet{Feldman2004}.}\label{fig:close}
\end{figure*}

The inferred hydrogen
abundance at Gale crater changes little between the raw data and the reconstruction (lower panels in figure~\ref{fig:close}), being around 8.0 $\pm$ 0.4 wt\% WEH in both
cases.  However we find a gradient in the higher resolution reconstruction, with hydration increasing towards the
south-east.  It is worth noting that these results are discrepant with those obtained by DAN, which suggest an average abundance of 2.1 to 2.7
wt\% WEH \citep{Mitrofanov2014} and those from the High Energy Neutron Detector (HEND) instrument in the Mars Odyssey Gamma Ray Spectrometer suite with 5 wt\% WEH \citep{Mitrofanov2016}.  It has been suggested that
this discrepancy may be the result of the greatly differing sizes of the response functions of the ground
and space based detectors (3 m for DAN, 300 km for HEND and 520 km for MONS) and the presence of the rover in a locally dry
region \citep{Mitrofanov2014}. However, if this were the correct explanation we may expect that improving the resolution would go
some way towards alleviating this discrepancy and we find no evidence of this in our reconstruction.  For this explanation to hold, the variation in abundance must be on a scale smaller than the scales accessible to the
reconstruction (i.e. 290 km).  Alternatively, the differing mean depths from which the detected neutrons originate may be
responsible: DAN is sensitive only to the composition of the top 60 cm of the surface \citep{Mitrofanov2014}, whereas the
orbital detectors see up to 1 m into the surface, depending on the soil water abundance.  If the deeper regions are more hydrated than the surface then we would expect the orbital detectors to measure larger WEH values than DAN.


\section{Possible sites of remnant hydration from periods of high obliquity}\label{sec:small}
\begin{figure*}[]
	\begin{center}
	\includegraphics[width=0.9\textwidth]{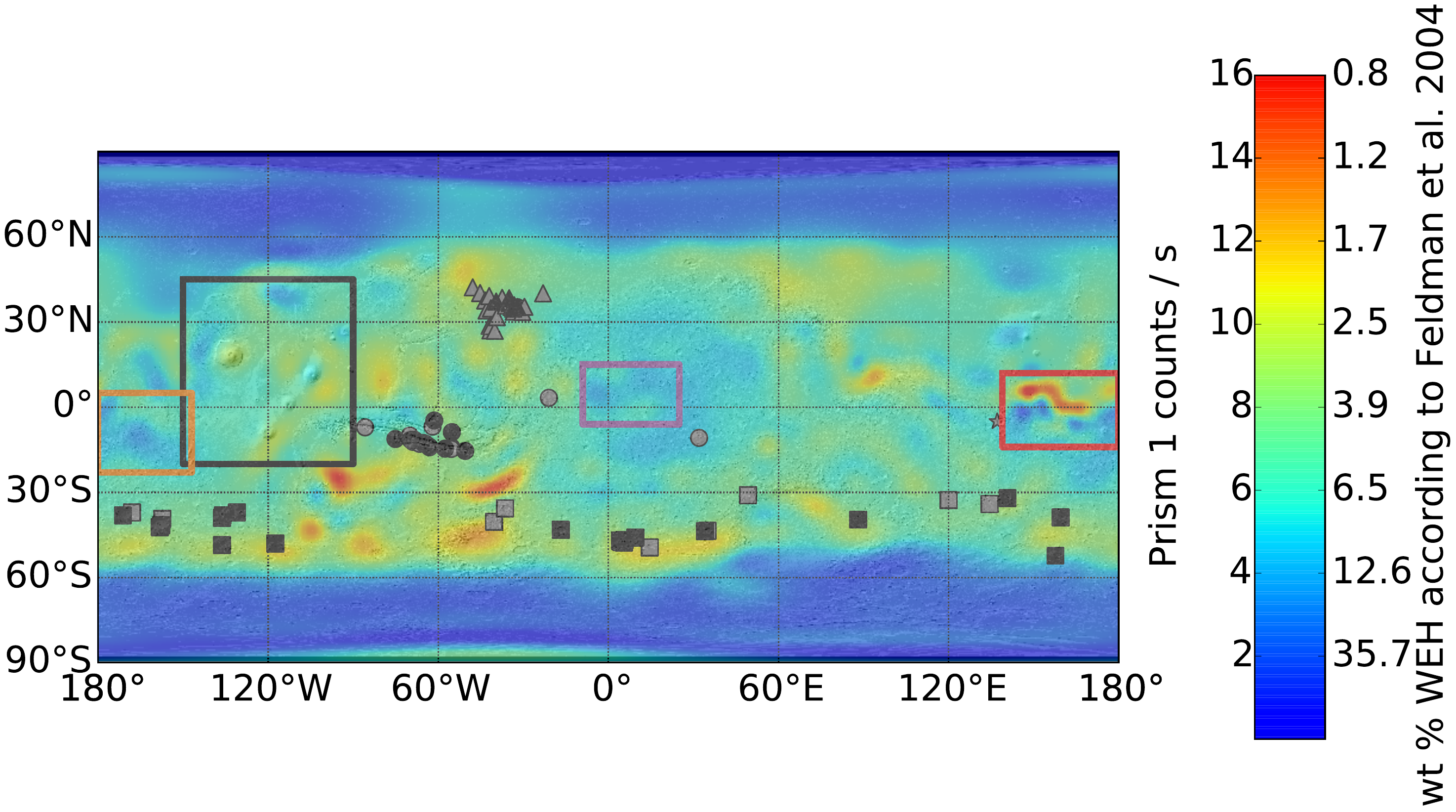}
	\end{center}
\caption{Global MONS reconstruction outlining the regions examined in later sections. The Tharsis Montes, examined in section~\ref{sec:tharsis}, are outlined in black, Elysium planitia and part of the Medusae Fossae Formation, in red, the remaining section of the Medusae Fossae Formation, in orange, and Meridiani Planum in violet. The locations of both confirmed (black) and candidate (grey) RSL are also shown as identified in \citet{McEwen2011} (squares),  \citet{Stillman2014} (circles) \citet{Stillman2016} (triangles) and \citet{Dundas2015} (star).}\label{fig:1}
\end{figure*}

We showed, in section~\ref{sec:res}, that pixon reconstruction of the MONS data yields a near two-fold improvement in linear spatial resolution and a 50\% increase in dynamic range. Here, we will use these results to examine a series of science targets  that have sizes close to or smaller than the MONS PSF
and are suggested, or believed, to contain contemporary water ice or hydrated minerals.  The selected sites are outlined in figure~\ref{fig:1}.

In this section we will focus on a few locations that have been proposed to contain water in the equatorial regions of Mars deposited during periods of high orbital obliquity and use the results of the previous section to infer their present hydration.

\subsection{The Medusae Fossae Formation: ice or dry, porous rock?}

\begin{figure*}[]
	\begin{center}
	\includegraphics[width=\textwidth]{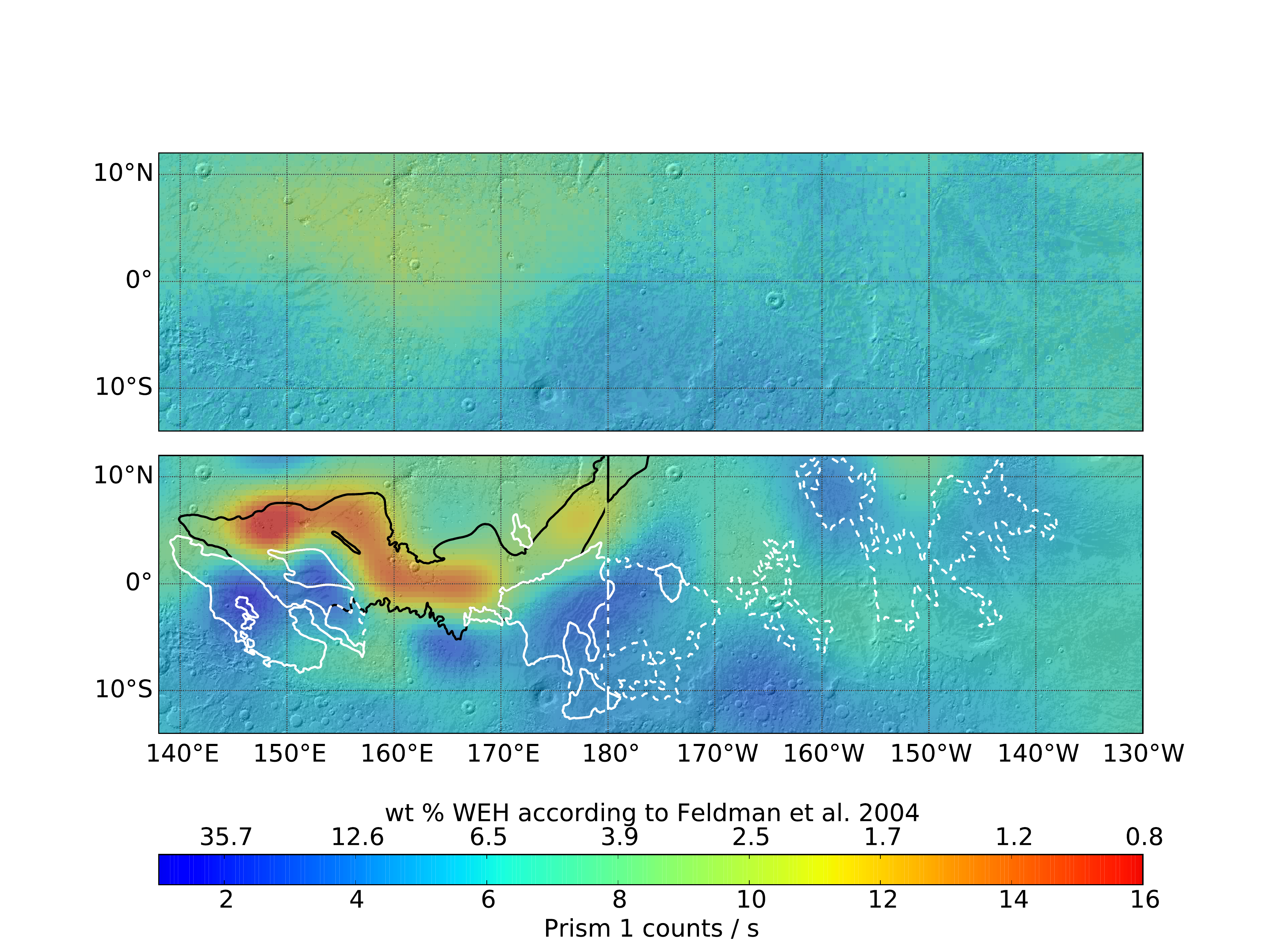}
	\end{center}
\caption{({\it Top}) Raw MONS data and ({\it Bottom}) pixon reconstruction around the MFF. A MOLA shaded relief map is underlayed.  The contours outline the young undivided channel material (black) and middle and lower members of the MFF (white solid and dashed, respectively) identified in \citet{Scott86}.}\label{fig:mff}
\end{figure*}

The topography and MONS reconstruction of the MFF are shown in figure~\ref{fig:mff}.  The western lobes, Aeolis, Zephyria and Lucus Plana, of the MFF are greatly enriched in hydrogen, with $>$10 wt. \% WEH.  At Aeolis Planum the neutron data imply a WEH abundance $>$ 40 wt. \%.  This is too great an abundance to be explained as hydrous silicates, which have WEH abundances of 10-20~wt.~\% \citep{Feldman2004}, so must be buried water ice or hydrated salts.  WEH abundances above $\sim 26$~wt.~\% cannot be accommodated as ice within soil pore spaces, so if ice is present at the MFF it must be present as bulk ice \citep{Feldman2011}.  However, the eastern lobes of the MFF contain little, if any, increased hydration. Geologically, the western lobes are associated with the lower members of the deposit, whereas the eastern lobes corresponds to upper and middle members only \citep{Scott86,Greeley87}.  The region of enhanced hydration at 10\degrees\,S, 165\degrees\,W is not coincident with the MFF, but does lie on the dichotomy boundary (the line separating the Martian northern lowlands and southern uplands).

At present water ice is unstable at any depth near the Martian equator \citep{Schorghofer2005,Schorghofer2007}.  However, Martian orbital obliquity has varied greatly over the past 20~Ma \citep{Laskar2004}.  During high obliquity periods the polar water ice caps become unstable and ice may be deposited down to equatorial locations \citep{Forget2006}.  Protection of this deposited ice by dust, perhaps containing cemented duricrust layers, could lead to its continued presence today \citep{Feldman2011,Wang2011}.

Our result is consistent with the lower member of the MFF containing ice rich material, which lends weight to the theory that, at least part of, the MFF is a polar layered-like deposit. Salt hydrates could also explain the observations as they have up to 50~wt.~\% WEH, but their stability in the Martian regolith has not been demonstrated \citep{Bish2003}. No detections of hydrated minerals within the MFF are reported in either CRISM or OMEGA data sets \citep{Carter2013}.

\subsection{Tropical mountain glaciers at the Tharsis Montes}\label{sec:tharsis}

\begin{figure*}[]
	\begin{center}
	\includegraphics[width=\textwidth]{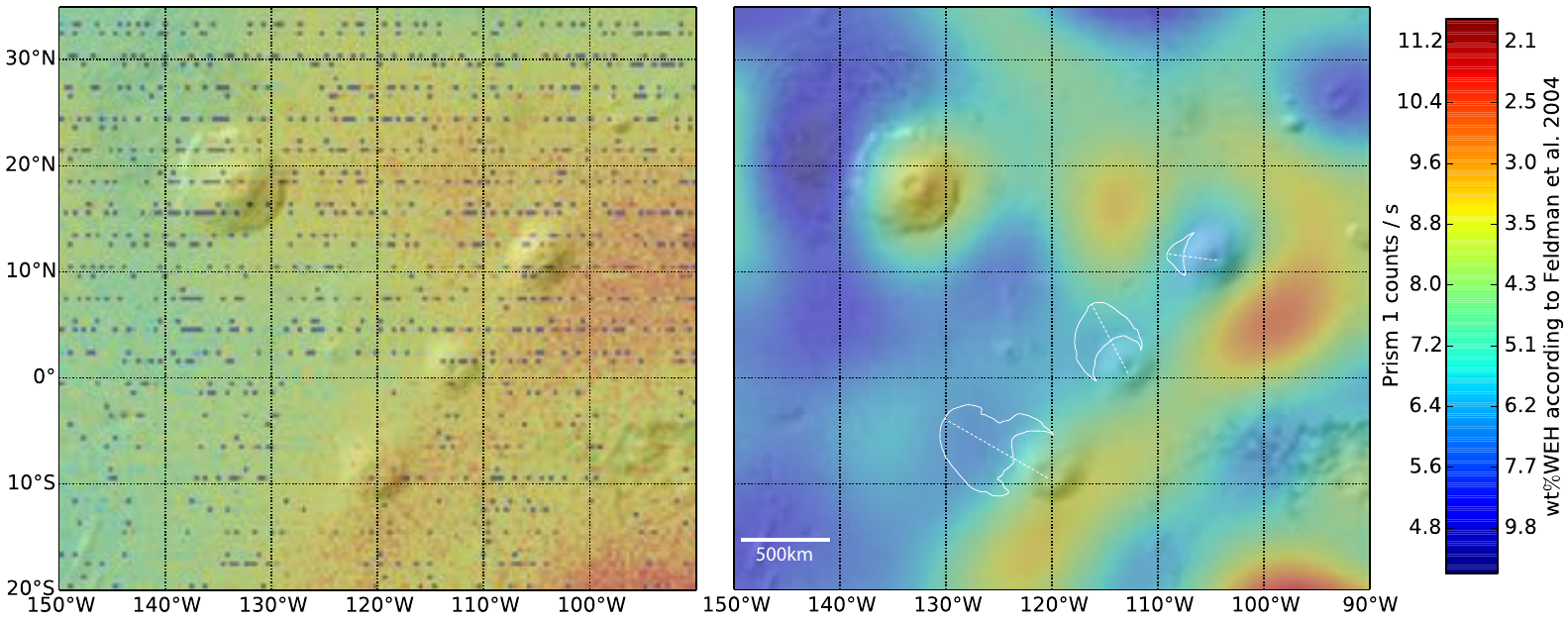}
	\end{center}
\caption{{\bf Left}: Gridded MONS prism-1 data at Tharsis. {\bf Right}: Pixon reconstruction of the MONS prism-1 data at Tharsis, showing the WEH conversion of \citet{Feldman2004}. Underlaid is a MOLA shaded relief map. The white contours show fan shaped deposits on the slopes of the Tharsis Montes (and are based on those in \citet{Fastook2008}. }\label{fig:tharsis}
\end{figure*}

The Tharsis region, shown with a black rectangle in figure~\ref{fig:1}, is presented in more
detail in figure~\ref{fig:tharsis}. In our pixon reconstruction in the right panel of figure~\ref{fig:tharsis}, the line of the
Tharsis Montes is seen to separate regions of higher and lower WEH abundances, where the north-west side of the line is enhanced in hydrogen with abundances $>$7 wt\% WEH on the north-west flanks of the Tharsis Montes. This hydrogen concentration is greater than that in the raw data (figure~\ref{fig:tharsis} left panel) and than in Elphic et al. (2005a). The area with the highest WEH concentration is north-west of Olympus Mons (at 18.65\degree\ N, 134\degree\ W) and coincides with the Lycus Sulci region where the WEH content reaches 11 wt\%.

In addition to these hydrogen enhancements, \citet{Forget2006} suggest that ice should condense west of Elysium Mons (25.02\degree\ N, 147.21\degree\ E) during periods of high orbital obliquity. In figure~\ref{fig:1} we also find a decreased epithermal neutron count rate, implying enhanced WEH, at this location. \citet{Forget2006} also note that if the source for the atmospheric water during periods of high obliquity were not the north polar cap alone but included an additional, hypothetical, south polar water cap, then ice would be deposited in the Eastern Hellas basin (110\degree E, 40\degree S).
However, we see no evidence for this in figure~\ref{fig:1}.

That we see a peak in the count rate in the reconstruction so closely coincident with the position of Olympus Mons in MOLA topography (figure~\ref{fig:tharsis}) strengthens our faith in the other features shown in the reconstruction. The region of enhanced hydrogen content around Ascraeus Mons (107\degree W, 12\degree N) is larger than the glacial deposit identified by \citet{Parsons2005}. This may be an interesting result, although establishing its significance will require determining the precise spatial resolution of the reconstruction at this location.

\section{Other locations with evidence of present water}\label{sec:other}

In this section we will examine in our pixon reconstruction of the MONS data several locations that are suggested on the basis of other spectral or imaging data to show evidence of past or present hydration.  The MONS reconstruction will be used to reveal whether the top metre of the surface is hydrated at these locations today.

\subsection{Elysium Planitia: ice or lava?}

The Cerberus fossae are a 1600 km long set of parallel fissures at Elysium planitia, in equatorial Mars.  The fissures stretch from 154.43{\degrees}E, 16.16\degrees N to 174.72\degrees E, 6.23\degrees N.  The fault is believed to be related to the Elysium Montes, located to the northwest, and there is morphological evidence that it has been the source of both water and lava floods \citep{Berman2002,Burr2002,Head2003} within the last 2-10 Myr.  It had been assumed that the water had sublimated away, leaving only the fluvial channels of Athabasca and Marte Valles as evidence of the water flows.  However, \citet{Murray2005} identified, in Mars Express High Resolution Stereo Camera (HRSC) images, plate-like features that they argued must be the remains of a buried frozen sea 800$\times$900 km across and up to 45 m deep, centred at 5\degrees\,N, 150\degrees\,E.  Their argument is based on three observations: the geomorphological similarities of the HRSC images with those taken above Antartica showing the break up of pack-ice, discrepancies between the ages of the plates and inter-plate regions, and the reduction in the volume of craters that have been filled by the flow. 



The epithermal neutron flux is sensitive only to the top metre of the surface, which, if the frozen sea were present, may consist of sublimation lag or pyroclastic deposits from subsequent eruptions from the Cerberus fossae.  Nonetheless, if the outflow occurred within the last 10 Myr we may still expect to see excess hydration due to partial protection of the water by a dust cover lag, perhaps containing hydrated mineral layers that provide a barrier protecting a deeper, higher-humidity environment \citep{Feldman2011,Wang2011}. 

The locally adaptive pixon reconstruction of the region around the proposed, buried sea is shown in figure~\ref{fig:mff}.  The location of the water ice sea, 5\degrees\,N, 150\degrees\,E, corresponds to one of the locations with the highest epithermal neutron flux on the entire surface of Mars.  This suggests that the top few tens of centimetres of soil, in this region, are unusually dry with $<1$ wt. \% WEH.  The dry feature in the reconstruction extends beyond the region identified by \citet{Murray2005} and covers much of the smooth plains that are believed to be young basaltic lavas from Cerberus fossae (outline with a black contour in figure~\ref{fig:mff}).  Additionally, \citet{Boynton2007}, using Mars Odyssey Gamma Ray Spectrometer data, find this region to be enhanced in Fe.  Taken together, these result suggests that the plate-like features at Elysium Planitia are unlikely to be a buried water ice sea but are, instead, young, Fe-rich, volatile-poor basalts.  Although no enhancement in the near-subsurface hydrogen content is seen in this region the very high epithermal neutron flux suggests that it is compositionally unique.  Previous work using MONS data has considered only the effects of H and CO$_{\rm 2}$ on the epithermal neutron flux. To understand the effect of changing the abundances of other elements, such as iron, would require new neutron transport simulations. 

The results of section~\ref{sec:errors} indicate a systematic error of up to 1.5 counts per second and a random error of 0.6 counts per second in this region.  These errors are amongst the highest on the surface, however even in the worst case scenario that the random and systematic errors were in the same direction southern Elysium planitia remains an exceptionally dry region.  Thus, our conclusions will not be affected, qualitatively, by errors in the reconstruction technique.

\subsection{Meridiani Planum}
\begin{figure}[]
	\begin{center}
	\includegraphics[width=\columnwidth]{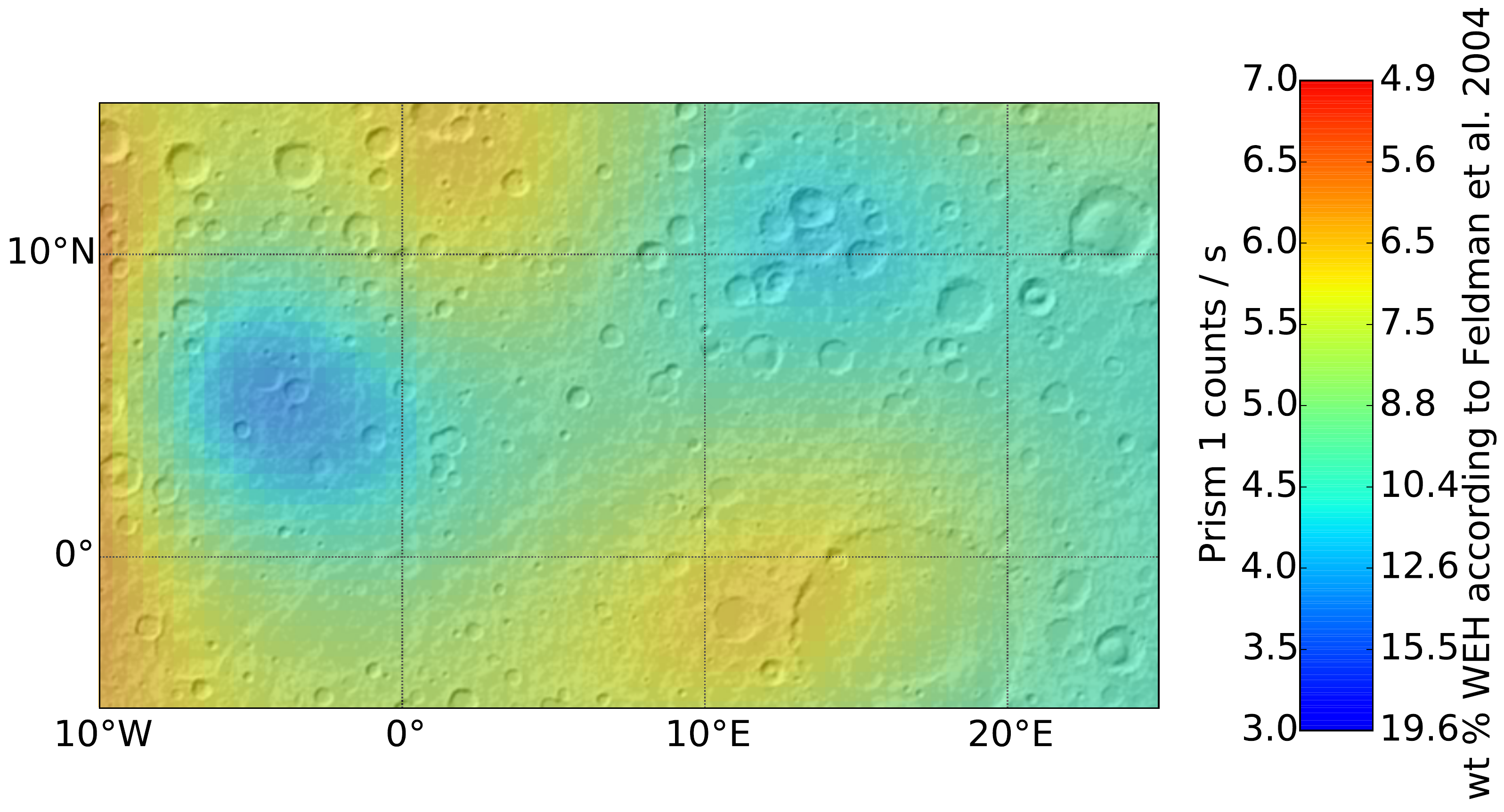}
	\end{center}
\caption{Locally adaptive pixon reconstruction of the MONS epithermal data around Meridiani Planum.}\label{fig:Meridiani}
\end{figure}

Meridiani Planum is a plain centred at 0.2\degree\,N, 2.5\degree\,W and near the dichotomy boundary.  The plains became an area of intense interest after they were identified as being one of the few locations containing, and the only large deposit of, crystalline, grey hematite on Mars \citep{Christensen2001}.  The favoured explanation for the origin of this mineral is chemical precipitation from aqueous Fe-rich fluids.  This process requires the presence of liquid water, on the surface of Mars, for substantial periods of time, early in its history \citep{Christensen2001}. For this reason the region was selected as the landing site for NASA's Mars Exploration Rover, Opportunity. 

The grey hematite deposits are centred on 4\degree\,W, 2\degree\,S. Comparison with figure~\ref{fig:Meridiani} shows this area is not enriched in H in the MONS reconstruction.  However, an area of the Planum $\sim$600 km to the north, at 5\degree\,W, 4\degree\,N, does show hydrogen enrichment with $\sim$15 wt. \% WEH, consistent with the presence of hydrated minerals. This region was shown, using the OMEGA visible-near infrared hyperspectral imager, to contain hydrated minerals including kieserite, gypsum, and polyhydrated sulfates \citep{Gendrin2005}.  Why the hematite deposits appear less hydrogen rich than hydrated mineral deposits to their north is not clear.

\subsection{Recurring slope lineae}\label{sec:rsl_int}
Recurring slope lineae are narrow, dark markings that appear and grow on Martian steep slopes during warm seasons and fade in 
cold seasons.  They have been found at equatorial and mid-latitudes and were first identified using HiRISE images 
\citep{McEwen2011}. Contemporary liquid water activity on the surface of Mars has recently been confirmed at sites showing 
recurring slope lineae, using infra-red spectral data from the Compact Reconnaissance Imaging Spectrometer for Mars onboard the Mars 
Reconnaissance Orbiter \citep{Ojha2015}.  

The origin and recharge mechanism of these liquid brines detected in recurring slope lineae (RSL) has important implications for the current Martian water cycle and budget. However the exact mechanism that recharges the RSL, which may vary across the surface, remains poorly understood. At least three possible scenarios exist:
feeding from shallow briny aquifers \citep{McEwen2011,Chevrier2012}; melting of shallow ice during warm seasons \citep{McEwen2014}, which provides good temporal agreement with RSL formation but may not be able to provide suffient water over long periods \citep{Stillman2014,Grimm2014}; and absorption of atmospheric water vapour by deliquescent salts.  The first two of these hypotheses require some subsurface store of water but the third does not.  Additionally, a dry hypothesis involving granular flow is possible \citep{McEwen2011}.

To distinguish between these hypotheses, we use the 
results of our locally adaptive pixon reconstruction of the MONS epithermal neutron data.

\subsubsection{Correlation of RSL position with H abundance}

\begin{figure}[]
	\begin{center}
	\includegraphics[width=\columnwidth]{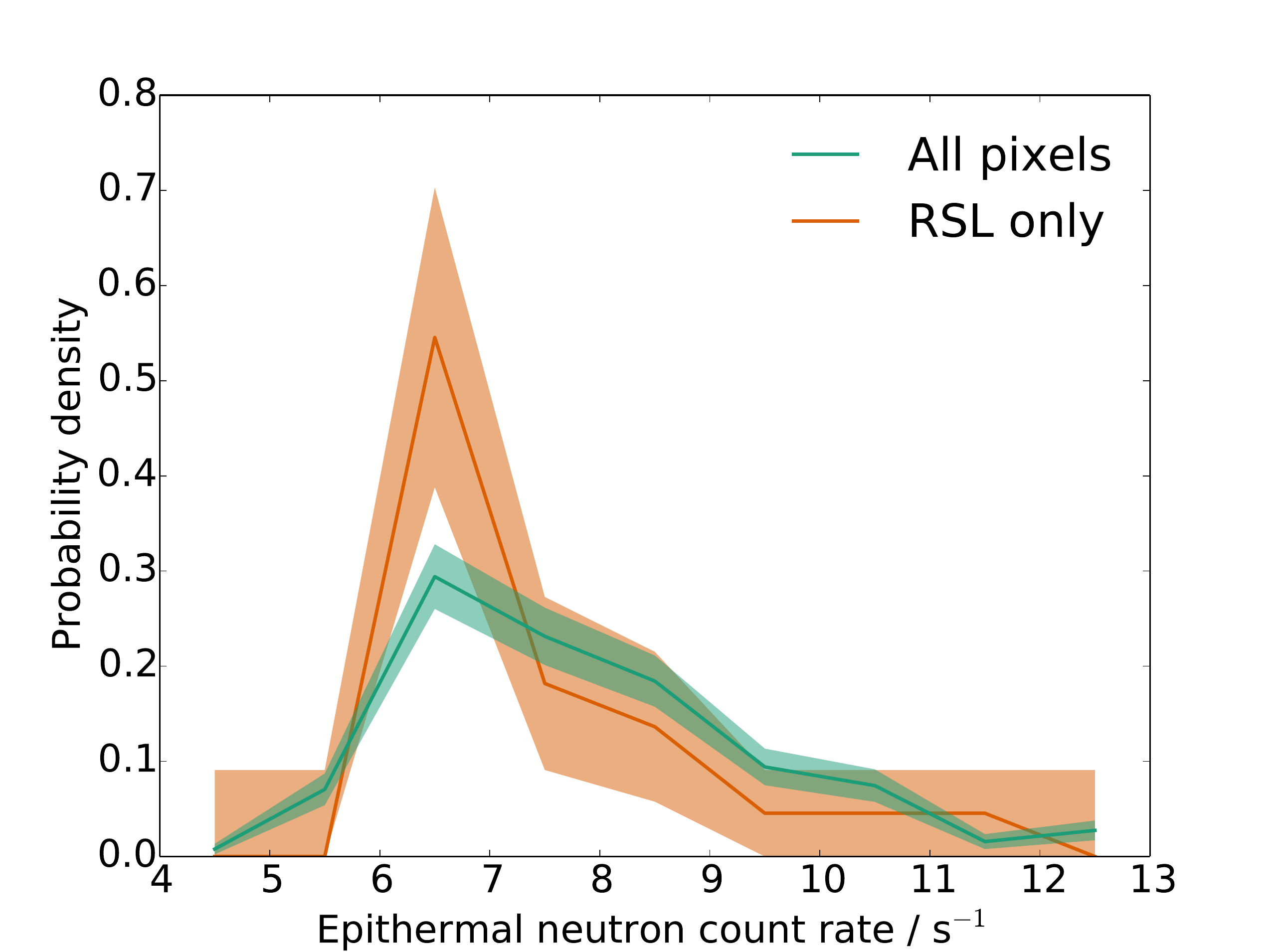}
	\end{center}
\caption{The probability density function of the count rate between 53\degree S and 30\degree S and just those pixels containing RSL. The translucent regions indicate 1$\sigma$ errors.}\label{fig:pdf}
\end{figure}

Comparison of the high resolution hydrogen map with the locations of RSL shows no positive 
correlation (figure~\ref{fig:1}).  To test this comparison and find whether RSL occur in special regions in the WEH map we will examine the probability density function of the count rate. Figure~\ref{fig:pdf} shows the probability density function of the 
reconstructed count rate in the southern mid-latitudes (SML), along with the probability distribution of the count rate restricted to 
sites within the SML containing RSL.  We compared these two distributions using a two-sample Kolmogorov-Smirnov test \citep{NR}, which 
measures the maximum difference between the cumulative distribution functions of two samples. This tests the null hypothesis that the data sets are drawn from the same distribution. Large values of the test statistic 
imply that the two samples are not drawn from the same underlying distribution.  To carry out this calculation the data were repixelated more coarsely so that adjacent pixels are effectively uncorrelated.  We find a Kolmogorov-Smirnov test statistic of 0.2 and 
corresponding p-value of 0.2 (i.e. test statistics at least this extreme are expected 20\% of the time).  This is not sufficient to imply that the count-rates (and consequently water abundances) where 
RSL are present are different from those locations where RSL are absent.  In turn, this result implies that the occurrence of RSL is independent of the WEH of the top few 10s of cm of soil on large scales. It should be noted that this analysis assumes that the distribution of discovered RSL reflects their true distribution across the surface in an unbiased way.  If there are strong selection effects that correlate with WEH abundance then the results may change, for example if the low albedo regions in which RSL are found also have anomalous hydrogen abundances.

That WEH abundance in the soil is not correlated with RSL activity can be explained either by reference to the resolution of the neutron data, even at 290 km the resolution is too coarse and the regions containing RSL are blurred into those without, or we must conclude that there are no sizeable stores of subsurface water close to the RSL thus their formation and renewal is not driven by subsurface aquifers or water ice.  Water buried beneath the level accessible to the MONS data (i.e. deeper than about 1 m) could not act as a source for the RSL because thermal inertia prevents ice buried beneath 20 cm from ever being melted \citep{Stillman2014}.  It was shown in section~\ref{sec:min_size} that water deposits smaller than 60 km in radius do not produces statistically significant decreases in epithermal neutron count rate, so RSL sources smaller than this are not ruled out by this work.

If the RSL water were sourced from a subsurface reservoir too small to be seen in MONS data then we may expect that improving the resolution would go some way towards alleviating this discrepancy. However we find no evidence of a systematic increase in water abundance at the RSL with reconstruction.  If the water source for the RSL is not subsurface then it must be either atmospheric or the sublimation of deeply buried ice deposits. But, both of these mechanisms require that the 
deliquescence of chlorate and perchlorate salts (the presence of which has been confirmed by CRISM and in-situ measurements 
\citep{Hecht2009,Glavin2013,Ojha2015}) produces sufficient quantities of liquid brines to darken the soil and produce the RSL.

\section{Conclusions}\label{sec:conc}

We have investigated several regions of interest across the Martian surface that have previously been hypothesised to contain water or hydrated minerals. A consequence of blurring by the detector footprint is that earlier analyses of the neutron count rates had led to underestimation of the dynamic range of hydrogen abundance in regions with local variation.

At low and equatorial latitudes we found evidence, in the reconstruction of the MONS data, for buried water ice in the Medusae Fossae Formation and on the western slopes of the Tharsis Montes and Elysium Mons. This supports the hypothesis that Mars in the recent geological past rotated on an axis highly inclined to its current one.  On the other hand, the Cerberus fossae region containing plate-like features, suggested to be a buried water ice sea, was found to be exceptionally dry.  Although this region seems not to contain hydrogen within the top few tens of cm of the surface, its unusually high epithermal neutron flux is consistent with the area being unusually high in Fe and low in H.   Further work using Monte Carlo particle transport codes is required to learn more about the composition of this feature. 

Recurring slope lineae are transient features recently discovered on Martian steep slopes during warm seasons. Their narrow, dark shapes resemble water courses -- and the presence of liquid water has recently been spectroscopically confirmed near some RSL sites. Yet, if this is what they are, then their source remains unknown. We find that the hydrogen abundance, on scales of $\sim$290~km, at RSL sites is statistically no different from anywhere else at similar latitudes.  This implies that RSL are not supplied by large ($>$~160~km diameter), near-surface briny aquifers. Preferred hypotheses remain for RSL to either be fed by water vapour, either from sublimating deeply buried water ice or from the atmosphere, or by small ($<$~160~km diameter) or deeply ($>$~1~m) buried aquifers.

\section*{Acknowledgement}
JTW is supported by the Science and Technology
Facilities Council [grant number ST/K501979/1] and Cosmiway [grant number GA 267291]. VRW is supported by the Science and Technology
Facilities Council [grant number ST/L00075X/1].
RJM is supported by a Royal Society University Research Fellowship.
This work used the DiRAC Data Centric system at Durham University,
operated by the Institute for Computational Cosmology on behalf of the
STFC DiRAC HPC Facility (www.dirac.ac.uk). This equipment was funded
by BIS National E-infrastructure capital grant ST/K00042X/1, STFC
capital grant ST/H008519/1, and STFC DiRAC Operations grant
ST/K003267/1 and Durham University. DiRAC is part of the UK national
E-Infrastructure.

\appendix

\section{Convolution with an azimuthally symmetric bandlimited function}\label{sec:fastConv}
The convolution between a circularly symmetric (or zonal) function (i.e. one that depends only on arc-length from a particular point, taken to be the north pole), $G$, with an arbitrary function, $f$, defined on the sphere can be written:
\begin{equation}\label{eqn:conv}
(G*f)(\bm{\xi}) = \int_{\mathbb{S}^2} G(\bm{\xi} \cdot \bm{\xi'})f(\bm{\xi'}) d\bm{\xi'}
\end{equation}
i.e. an integral over the 2-sphere.

The spherical harmonic (SH) representation of $G$ is
\begin{equation}
G(\bm{\xi}) = \sum_{l=0}^{\infty}\sum_{m=-l}^{l}g_l^mY_l^m(\bm{\xi}).
\end{equation}
As $G$ is circular, $a_{lm}=0$ for $m\neq 0$ thus
\begin{equation}
\begin{split}	
G(\bm{\xi}) &= \sum_{l=0}^{\infty}g_l^0Y_0^l(\bm{\xi})\\
G(\bm{\xi}) &= \sum_{l=0}^{\infty}\sqrt{\frac{2l+1}{4\pi}}g_l^0P_l(\bm{\xi}),
\end{split}
\end{equation}
where $P_l$ is the $l$th Legendre function. substituting this back into equation~(\ref{eqn:conv}) gives
\begin{equation}
\begin{split}
(G*f)(\bm{\xi}) &= \sum_{l=0}^{\infty}\int_{S^2} \sqrt{\frac{2l+1}{4\pi}} g_l^0P_l(\bm{\xi} \cdot \bm{\xi'})f(\bm{\xi'}) d\bm{\xi'} \\
&= \sum_{l=0}^{\infty}\sum_{m=-l}^{l} \sqrt{\frac{4\pi}{2l+1}} g_l^0 Y_l^m(\bm{\xi}) \int_{S^2} Y_l^m(\bm{\xi'})f(\bm{\xi'}) d\bm{\xi'}\\
&= \sum_{l=0}^{\infty}\sum_{m=-l}^{l} \sqrt{\frac{4\pi}{2l+1}} g_l^0f_l^m Y_l^m(\bm{\xi}),
\end{split}
\end{equation}
which can be seen to be a simple scaling of the SH coefficients of the function $f$.  These scaling 
factors can be calculated empirically.  In the case of convolution with a Gaussian function, of width $\sigma$, the 
appropriate SH coefficients can be calculated by solving for the fundamental solution of the heat 
equation, on the sphere. We find that the appropriate solution is
\begin{equation}
g_l^m = \sqrt{\frac{2l+1}{4\pi}}e^{\frac{-l(l+1)\sigma^2}{2}}.
\end{equation}

This algorithm has a time complexity of $\mathcal{O}(l_{\rm max}^2)$, where $l_{\rm max}$ is the maximum order used in our spherical harmonic decompositions, so in practice will be dominated by the time taken to perform the spherical harmonic transforms of the function being convolved, which in our implementation has $\mathcal{O}(l_{\rm max}^3)$ complexity.  This method is therefore $\mathcal{O}(l_{\rm max})$ faster than the `fast' method described in the previous section and $\mathcal{O}(l_{\rm max}^2)$ faster than the brute force method.  As, in this work, $l_{\rm max}$ will typically be $\mathcal{O}(10^2)$, care will be taken to formulate the deconvolution techniques such that all convolutions are with azimuthally symmetric functions.

\section{Integration of functions defined in pixel space}\label{sec:pixSpace}
It should be noted that the integrations in this and the previous section are done over 
solid angle rather than pixels. To convert from one coordinate system to the other, note that
\begin{align}\label{eqn:dxdohm}
\int d{\bf x} &= N_{\rm{pix}} \\
&= \int \frac{\partial{\bf x}}{\partial\bm{\Omega}}\,d\bm{\Omega} \\
&= \int \frac{1}{a\left(\bm{\Omega}\right)}\,d\bm{\Omega},
\end{align}
where $a\left(\bm{\Omega}\right)$ is the area of the pixel at position $\bm{\Omega}$, which is given by

\begin{equation}
a\left(\bm{\Omega}\right) = a\left(\theta, \phi\right) = \Delta\phi\left(\cos\left(\theta-\frac{\Delta\theta}{2}\right) - \cos\left(\theta + \frac{\Delta\theta}{2}\right)\right).
\end{equation}
$\Delta\phi$ is the width of the pixels in longitude, i.e. $\Delta\phi = \frac{2\pi}{X}$, where $X$ is the number of pixels in the longitudinal direction, similarly, $\Delta\theta = \frac{\pi}{Y}$. 




\bibliographystyle{elsarticle-harv}
\bibliography{refs.bib}

\end{document}